\documentclass[prd,aps,twocolumn,amsmath,amssymb,nofootinbib,preprintnumbers]
{revtex4}

\usepackage{graphicx}
\usepackage{dcolumn}
\usepackage{bm}
\usepackage{amsmath}
\usepackage{amsfonts}
\usepackage{subfigure} 

\newcommand{\remove}[1]{}

\def\ls{\mathrel{\lower4pt\vbox{\lineskip=0pt\baselineskip=0pt
           \hbox{$<$}\hbox{$\sim$}}}}
\def\gs{\mathrel{\lower4pt\vbox{\lineskip=0pt\baselineskip=0pt
           \hbox{$>$}\hbox{$\sim$}}}}
\def\drawbox#1#2{\hrule height#2pt

\hbox{\vrule width#2pt height#1pt \kern#1pt
              \vrule width#2pt}
              \hrule height#2pt}

\def\Asym#1#2{\vcenter{\vbox{\drawbox{#1}{#2}
              \kern-#2pt       
              \drawbox{#1}{#2}}}}

\def\QBL{Q_{\scriptscriptstyle B-L}}

\newcommand{\beq}{\begin{equation}}
\newcommand{\eeq}{\end{equation}}

\begin{document}

%
\title{Prospects for indirect detection of sneutrino dark matter with IceCube}

\author{Rouzbeh Allahverdi$^{1}$}
\author{Sascha Bornhauser$^{1}$}
\author{Bhaskar Dutta$^{2}$}
\author{Katherine Richardson-McDaniel$^{1}$}

\affiliation{$^{1}$~Department of Physics \& Astronomy, University of New Mexico, Albuquerque, NM 87131, USA \\
$^{2}$~Department of Physics, Texas A\&M University, College Station, TX 77843-4242, USA }


\begin{abstract}

We investigate the prospects for indirect detection of right-handed sneutrino dark matter at the IceCube neutrino telescope in a $U(1)_{B-L}$ extension of the MSSM. The capture and annihilation of sneutrinos inside the Sun reach equilibrium, and the flux of produced neutrinos is governed by the sneutrino-proton elastic scattering cross section, which has an upper bound of $8 \times 10^{-9}$ pb from the $Z^{\prime}$ mass limits in the $B-L$ model. Despite the absence of any spin-dependent contribution, the muon event rates predicted by this model can be detected at IceCube since sneutrinos mainly annihilate into leptonic final states by virtue of the fermion $B-L$ charges. These subsequently decay to neutrinos with 100\% efficiency. The Earth muon event rates are too small to be detected for the standard halo model irrespective of an enhanced sneutrino annihilation cross section that can explain the recent PAMELA data. For modified velocity distributions, the Earth muon events
  increase substantially and can be greater than the IceCube detection threshold of 12 events $\mathrm{km}^{-2}$ $\mathrm{yr}^{-1}$. However, this only leads to a mild increase of about 30\% for the Sun muon events. The number of muon events from the Sun can be as large as roughly 100 events $\mathrm{km}^{-2}$ $\mathrm{yr}^{-1}$ for this model.

\medskip


\end{abstract}
MIFP-09-25 \\ September 24, 2009
\maketitle
\section{Introduction}

There are various lines of evidence supporting the existence of dark matter in the universe, but its identity remains a major problem the solution to which likely rests at the interface of particle physics and cosmology. It is well established that particle physics can explain dark matter in the form of weakly interacting massive particles (WIMPs)~\cite{WIMP}. In the standard scenario the dark matter relic abundance, as precisely measured by cosmic microwave background (CMB) experiments~\cite{WMAP5} is determined from the thermal freeze out of dark matter annihilation in the early universe. There are currently major experimental efforts for direct and indirect detection of dark matter particles. Indirect detection investigates annihilation of dark matter to various final states (photons, anti-particles, neutrinos) through astrophysical observations, while direct detection probes the scattering of the dark matter particle off nuclei inside dark matter detectors.

Supersymmetry is a front-runner candidate to address the hierarchy problem of the standard model (SM). The minimal supersymmetric standard model (MSSM) has become the focus of major theoretical and experimental activities for the past two decades. It has a natural dark matter candidate, namely the lightest supersymmetric particle (LSP), which can have the correct thermal relic abundance~\cite{EHNOS}.
It is also believed that there are gauge symmetries beyond those of the SM. A minimal extension of the SM gauge group, motivated by the nonzero neutrino masses, includes a gauged $U(1)_{B-L}$ gauge symmetry~\cite{mohapatra} ($B$ and $L$ are baryon and lepton number respectively). Anomaly cancellation then implies the existence of three right-handed (RH) neutrinos and allows us to write the Dirac and Majorana mass terms for the neutrinos to explain the light neutrino masses and mixings.

The $B-L$ extended MSSM
also provides new dark matter candidates: the lightest neutralino in the $B-L$ sector~\cite{khalil,ADRS} and the lightest RH sneutrino~\cite{inflation2}.
In this work we will focus on the sneutrino as the dark matter candidate\footnote{It is also possible to have successful inflation in the context of the $U(1)_{B-L}$ model~\cite{inflation1}. In this case the dark matter candidate (the RH sneutrino) can become a part of the inflaton field and thereby gives rise to a unified picture of dark matter, inflation and the origin of neutrino masses~\cite{inflation2}.}. The candidate is made stable by invoking a discrete $R$-parity, but in the context of a $B-L$ symmetry, a discrete matter parity can arise once the $U(1)_{B-L}$ is spontaneously broken~\cite{rparity}. The $B-L$ gauge interactions can yield the correct relic abundance of sneutrinos if the $U(1)_{B-L}$ is broken around the TeV scale.

Recently, it has been shown that it is possible to explain the positron excess observed in the PAMELA data~\cite{pameladata} in the context of a low scale $B-L$ extension of the MSSM~\cite{ADRS,Allahverdi:2009ae,Dutta:2009uf}. Due to a factor of 3 difference between the $B-L$ charges of the quarks and leptons, the anti-proton flux is naturally suppressed in this model in agreement with the PAMELA anti-proton data.
Furthermore, the $U(1)_{B-L}$ gauge coupling unifies with those of the SM symmetries, and the $B-L$ symmetry can be broken radiatively. The $B-L$ breaking around a TeV results in a $Z^{\prime}$ gauge boson with around a TeV mass that can be probed at the LHC along with the other new states of this model.

The RH sneutrino of this $B-L$ extended model can be detected when it elastically scatters off a nucleus. The sneutrino-proton scattering cross section is large enough to be probed in the ongoing and upcoming dark matter direct detection experiments~\cite{inflation2}. In addition, annihilation of sneutrinos at the present time produces LH neutrinos. It is interesting to investigate the possibility of indirect detection of sneutrino dark matter by using final state neutrinos in the IceCube neutrino telescope. This ongoing experiment plans to probe the neutrino flux arising from the annihilation of gravitationally trapped dark matter particles in the Sun and the Earth.
We will examine the status of the $U(1)_{B-L}$ model in two cases. In {\bf{case 1}}, the sneutrinos annihilate mostly into RH neutrinos that subsequently decay into LH neutrinos and the MSSM Higgs. In {\bf{case 2}}, the sneutrinos annihilate mostly into the lightest Higgs boson in the $B-L$ sector that decays into $\tau^+\tau^-$ and $b^+b^-$ quarks, which subsequently produce LH neutrinos via three-body decays. The recent PAMELA data~\cite{pameladata} can be explained in {\bf{case 2}}, where the final state taus give rise to the positron excess in the cosmic ray flux without producing a significant number of antiprotons~\cite{ADRS,Allahverdi:2009ae,Dutta:2009uf}. The large cross section required to explain the data arises from Sommerfeld enhancement~\cite{Sommerfeld} or from the non-thermal production of dark matter~\cite{Dutta:2009uf}.

Since the source of neutrinos are different in the two cases, two-body versus three-body decay, the energy spectrum of the neutrinos can be used to distinguish the cases. We will estimate the muon neutrino flux as well as the muon flux in both scenarios as a function of sneutrino mass. Since the Large Hadron Collider (LHC) is on the verge of producing physics results, it will enable us to measure the mass of the dark matter candidate. Therefore, using the LHC measurements and the IceCube results in tandem, we hope to discern the $B-L$ model. We will present predictions of this model using the standard dark matter halo model as well as the modified velocity distributions obtained in recent galaxy simulations.

This paper is organized as follows. In section II, we discuss the low scale $U(1)_{B-L}$ model. In section III, we give a general discussion of the indirect detection of sneutrino dark matter via neutrino final states. In section IV, we present our results and discuss the prospect of detection of sneutrino dark matter at IceCube in {\bf{case 1}} and {\bf{case 2}}. In section V, we show the results obtained for the modified velocity distributions. In section VI, we compare predictions for the sneutrino dark matter in the $U(1)_{B-L}$ model with those for the neutralino dark matter in the minimal supergravity model. Finally, we close by concluding in section VII.

\section{The $U(1)_{B-L}$ Model}

Since this $B-L$ is a local gauge symmetry, we have a new gauge boson $Z^{\prime}$ (and its supersymmetric partner). In the minimal model, we also have two new Higgs fields $H^{\prime}_1$ and $H^{\prime}_2$ (that are SM singlets) and their supersymmetric partners. The vacuum expectation values (VEVs) of these Higgs fields break the $B-L$ symmetry. We can write the superpotential of the model as follows (the boldface characters denote superfields)
\begin{equation} \label{sup}
W = W_{\rm MSSM} + W_{B-L} + y_D {\bf N} {\bf H_u} {\bf L}  \, ,
\end{equation}
where ${\bf H_u}$ and ${\bf L}$ are the superfields containing the Higgs field that gives mass to up-type quarks and the LH leptons respectively. For simplicity, we have omitted the family indices. The $W_{B-L}$ term contains ${\bf H^{\prime}_1},~{\bf H^{\prime}_2}$ and ${\bf N}$~\cite{Allahverdi:2009ae}. Its detailed form depends on the charge assignments of the new Higgs fields.

The $U(1)_{B-L}$ is broken by the VEV of $H^{\prime}_1$ and $H^{\prime}_2$, which we denote by $v^{\prime}_1$ and $v^{\prime}_2$ respectively. This results in a mass $m_{Z^{\prime}}$ for the $Z^{\prime}$ gauge boson. We have three physical Higgs fields $\phi,~\Phi$ (scalars) and ${\cal A}$ (a pseudo scalar). The masses of the Higgs fields follow $m^2_{\phi} < \cos^2 (2 \beta^{\prime}) m^2_{Z^{\prime}}$ (where ${\rm tan} \beta^{\prime} \equiv \langle H^{\prime}_2 \rangle/\langle H^{\prime}_1 \rangle$) and $m_{\Phi},~m_{\cal A} \sim m_{Z^\prime}$.

Various $B-L$ charge assignments are allowed by anomaly cancelation. We choose the charge assignment shown in Table 1. In this case $H^{\prime}_2$ couples to the RH neutrinos and gives rise to a Majorana mass upon spontaneous breakdown of the $U(1)_{B-L}$.
Choosing these Majorana masses in the $100~{\rm GeV}-1$ TeV range, we have three (dominantly RH) heavy neutrinos and three (dominantly LH) light neutrinos. The masses of the light neutrinos are obtained via the see-saw mechanism.

\begin{table}[h!]
\center
\begin{tabular}{|c||c|c|c|c|c|c|c|}\hline
{\rm Fields} & $Q$ & $Q^c$ & $L$ & $L^c$ & $H^{\prime}_1$ & $H^{\prime}_2$ \\ \hline
$\QBL$ & 1/6 & -1/6 & -1/2 & 1/2 &  1 & -1 \\ \hline
\end{tabular}
\caption{The $B-L$ charges of the fields for the minimal model. Here $Q$ and $L$ represent quarks and leptons respectively, while $H^{\prime}_1$ and $H^{\prime}_2$ are the two new Higgs fields. The MSSM Higgs fields have zero $B-L$ charges.}
\label{BLcharges-tbl}
\end{table}

A natural dark matter candidate in this model is the lightest sneutrino ${\widetilde N}$.
We note that it has fewer gauge interactions than other supersymmetric particles, and its mass receives the smallest contribution from the gaugino loops. Based on the dominant channel for sneutrino annihilation
we therefore consider the following two cases:
\begin{itemize}
{\item
{\bf Case 1}: A generic case where a solution to the positron excess observed by PAMELA is not sought. In this case the dominant annihilation channels are the $S$-wave processes ${\widetilde N} {\widetilde N} \rightarrow N N$ and ${\widetilde N}^* {\widetilde N}^* \rightarrow N^* N^*$ via $t$-channel exchange of ${\widetilde Z}^{\prime}$.
There are also ${\widetilde N} {\widetilde N}^* \rightarrow N N^*,~f {\bar f}$ annihilation modes via $s$-channel exchange of a $Z^{\prime}$ or $B-L$ Higgs fields, but these are $P$-wave suppressed and can be completely neglected (particularly at the present time). In this case the annihilation cross-section has the nominal value $\sim 3\times 10^{-26}$ cm$^3$/sec (dictated by thermal freeze out) at all times. The RH neutrinos produced from dark matter annihilation quickly decay to LH neutrinos and the MSSM Higgs.}
{\item
{\bf Case 2}: In this case the PAMELA puzzle is addressed via Sommerfeld enhancement of sneutrino annihilation at the present time~\cite{Allahverdi:2009ae}. In this part of the model parameter space the lightest $B-L$ Higgs $\phi$ is much lighter than the $Z^{\prime}$. The dominant annihilation channel is ${\widetilde N}^* {\widetilde N} \rightarrow \phi \phi$ via the $s$-channel exchange of the $\phi$ or $\Phi$, the $t$ or $u$-channel exchange of a ${\widetilde N}$, and the contact term $\vert {\widetilde N} \vert^2 \phi^2$. The interactions for these processes arise from the $D$-term part of the potential, and their strength is proportional to $m_{Z^{\prime}}$.
There are other $S$-wave processes with Higgs final states ${\widetilde N}^* {\widetilde N} \rightarrow \phi \Phi , ~ \phi {\cal A}, ~ \Phi \Phi,~{\cal A} {\cal A}$, but they are kinematically suppressed and/or forbidden. The annihilation modes ${\widetilde N} {\widetilde N} \rightarrow N N$ and ${\widetilde N}^* {\widetilde N}^* \rightarrow N^* N^*$ are also subdominant in this case. As in the previous case, annihilations to $f {\bar f}$ final states are $P$-wave suppressed and hence totally negligible. The cross section for annihilation to the $\phi \phi$ final state at the present time is required to be $3 \times 10^{-23}$ cm$^3$/sec in order to explain the PAMELA data. Sufficient Sommerfeld enhancement is obtained as a result of the attractive force between sneutrinos due to the $\phi$ exchange provided that the mass of $\phi$ is small ($<20$ GeV)\footnote{It is possible to invoke a non-thermal scenario where the sneutrinos are created from the decay of heavy moduli or 
 gravitinos~\cite{Dutta:2009uf}. In this case we do not need Sommerfeld enhancement to satisfy the PAMELA data, and the annihilation cross section will be large, $3 \times 10^{-23}$ cm$^3$/sec, at all times.}. The $\phi$ subsequently decays into fermion-antifermion pairs very quickly via a one-loop diagram, and it mostly produces $\tau^+\tau^-$ final states by virtue of the fermion $B-L$ charges~\cite{Allahverdi:2009ae}.}
\end{itemize}

The sneutrino-proton scattering cross section for this model
can be in the $10^{-11}-10^{-8}$ pb range for a reasonable choice of parameters that satisfy the
relic density constraint, cf.~\cite{inflation2,Allahverdi:2009ae}. This opens up the prospect
for direct detection with the help of the next generation of experiments~\cite{Direct}. The current upper bound for the spin-independent cross section is $4.6\times 10^{-8}-2 \times 10^{-7}$ pb for a dark matter mass of $60-1200$ GeV; this is just above the highest possible values for
our model\footnote{Since $B-L$ symmetry is vectorial, the spin-dependent cross section is zero in this model.}.

\section{Prospects for Indirect Detection at IceCube}

\subsection{The Neutrino Signal}
The $B-L$ model also shows great promise for indirect detection, and we focus in particular on the potential neutrino signal at the IceCube experiment. In {\bf{case 1}}, the sneutrinos annihilate to produce RH neutrinos that subsequently decay into a LH neutrino and a neutral Higgs boson\footnote{RH neutrino decay to a charged lepton and a charged Higgs is typically forbidden.}. We assume for most of this paper that the total LH neutrino flux branches into every neutrino flavor equally (see subsection~\ref{NeutrinoBranchingRatios} for a discussion). Assuming that the mass difference between the RH sneutrinos and RH neutrinos is small\footnote{This is the case when the soft supersymmetry breaking mass of the sneutrino is similar to or smaller than supersymmetry conserving Majorana mass of the (s)neutrino. A rather small soft mass term is motivated if the $B-L$ symmetry is to break radiatively and is needed to keep the lightest $B-L$ Higgs $\phi$ light as in {\bf {case 2}}~\cite{Allahverdi:2009ae}.}, the RH neutrinos are produced non-relativistically, and hence each LH neutrino and Higgs receives an energy equal to half of the sneutrino mass.

In {\bf{case 2}}, RH neutrinos constitute about 10\% of the annihilation final states. Two of the lightest $B-L$ Higgses $\phi$ compose the remaining $90\%$ of the branching fraction. This branching fraction is necessary to provide a high enough leptonic particle rate to fit the PAMELA data. As mentioned in the previous section we need $m_{\phi}<20$ GeV. For $4\,{\rm GeV}<m_\phi<20$ GeV, the final states are mostly taus (74\%) and b quarks (16\%), where the dominance of tau final states is a result of the fermion $B-L$ charges. The LH neutrinos in this case arise from the three-body decay of taus and bottom quarks. For $m_\phi<4$ GeV, we would have mostly muons and charm quarks.

Both the {\bf{case 1}} and {\bf{case 2}} scenarios of our model display a crucial signature difference when compared to the standard neutralino LSP in the MSSM. The energy distribution of the produced LH neutrinos from the RH neutrino decay is a delta function occurring at half of the sneutrino mass. Other annihilation channels in this model, as well as those available in the MSSM, produce additional neutrino signal via three-body decays such as  $\tau^- \rightarrow e^- \nu_{\tau}\bar{\nu_{e}}$. This difference opens up a significant possibility to differentiate between the $B-L$ model and the MSSM with the help of the differential energy spectrum of the detector event rates. This is discussed further in section~\ref{ModelResults}.

\subsection{Neutrino Flux} \label{Neutrino Flux}

Sneutrino annihilation in the Sun and the Earth produces an expected neutrino flux through IceCube. This flux is modeled by calculating the number of gravitationally captured sneutrinos and then considering the propagation and detection of the produced neutrinos. The number of captured dark matter particles as a function of time is governed by a differential equation the solution to which is
\begin{equation}
N(t) = \sqrt{\frac{C}{A}}\tanh{\sqrt{C A} t} \; ,
\end{equation}
where $C$ is the total capture rate and depends on both the total scattering cross sections off nucleons and $A$ is related to the annihilation cross section; see Ref.~\cite{Jungman:1995df} for details. The total rate of annihilation is given by
\begin{equation} \label{annihilationeq}
\Gamma_A = \frac{C}{2} \tanh^2{\left ( \frac{t}{\tau_{eq}} \right )} \;.
\end{equation}
The number of captured sneutrinos will saturate as long as the length of time for the process has exceeded the equilibration time, $\tau_{eq} \equiv (\sqrt{C A})^{-1}$.

In equilibrated systems, the rate of annihilation is entirely dominated by the capture rate C, $\Gamma_A \approx C/2$. We can explain equilibration in the $B-L$ model by considering some example cross sections. Since the age of the solar system is $4.5$ Gyr, for a $1$ TeV sneutrino with an annihilation cross section of $3 \times 10^{-23}$ cm$^3$/sec ($3 \times 10^{-26}$ $\mathrm{cm}^3/\mathrm{s}$), a spin-independent cross section $\sigma_{\mathrm{SI}}$ of at least $10^{-11}$ pb ($10^{-8}$ $\mathrm{pb}$) is needed to reach equilibration in the Sun. This assumes no spin-dependence as the $B-L$ model has none. The scattering cross section needed to achieve equilibration in the Earth is already excluded by direct detection bounds.

Alternatively we can fix the scattering cross section and place a limit on the annihilation cross section. In the $B-L$ model, the cross section for sneutrino-proton elastic scattering follows
\begin{equation} \label{sigma}
\sigma_{\rm SI} \propto \left({g_{B-L} Q_L \over m_{Z^{\prime}}}\right)^4 m^2_p ,
\end{equation}
where $g_{B-L}$ and $Q_L$ are the $U(1)_{B-L}$ gauge coupling and $B-L$ charge of leptons, respectively, and $m_p$ is the proton mass. The limits on the $Z^\prime$ mass from LEP and Tevatron are given by~\cite{tevatron,LEP},
\begin{equation} \label{Z'mass}
\frac{m_{Z^\prime}}{g_{B-L} Q_L} > 6 \; \mathrm{TeV} \;.
\end{equation}
This results
in an upper limit on $\sigma_{\mathrm{SI}}$ of $8 \times 10^{-9} \;  \mathrm{pb}$. Assuming this bound is realized, an annihilation cross section $\geq$ $4 \times 10^{-26} \mathrm{cm}^3/\mathrm{s}$ ($1 \times 10^{-18} \mathrm{cm}^3/\mathrm{s}$) needs to be achieved to reach equilibrium in the Sun (Earth). Note that we can always choose the $B-L$ gauge coupling and scale $B-L$ charges in accordance with anomaly cancelation such that $\sigma_{\rm SI}$ is saturated while obtaining the correct relic density for sneutrino dark matter. This is possible since a different combination of $g_{B-L}$ and $Q_{B-L}$ appears in the relic density calculation .
This is in contrast to the MSSM case where the SM gauge couplings and charges are fixed.

Since equilibrium is easily achieved in the Sun, the neutrino signal will depend solely on $C$, or equivalently $\sigma_{\mathrm{SI}}$, so the increased annihilation rate in {\bf {case 2}} of our model confers no advantage compared to typical MSSM cases for annihilation in the Sun. On the other hand, choosing reasonable values for either of the relevant cross sections demonstrates that equilibrium is nearly impossible to reach for the Earth without significant deviation from the assumptions made in~\cite{Jungman:1995df}. Consequently, the neutrino signal from the Earth will depend on both $C$ and $A$. Therefore one expects a much larger signal for {\bf {case 2}} as compared to either {\bf {case 1}} or the neutralino dark matter models~\cite{Perez}.

The annihilation of sneutrinos in the Sun and Earth yields neutrinos that can be detected by the IceCube experiment. IceCube can distinguish between neutrino signals from the Earth and Sun with the help of an angle cut.
This cut restricts the detection to an angle range of $90^{\circ}<\Theta<113^{\circ}$ in the case of the Sun, where $\Theta$ is the Earth zenith angle. One has to measure below the horizon to be able to distinguish the background of atmospheric neutrinos from the signal, and the Sun cannot be more than $23.5^{\circ}$ below the horizon at the South Pole \cite{Rizzo:2007zz,Ackermann:2005fr}. In the case of a search for a potential Earth signal one looks at a zenith angle of about $180^{\circ}$, i.e.,~directly to the core of the Earth~\cite{Rizzo:2007zz}.

Muon neutrinos create muons via charged current interactions in the detector. The qualitative behavior of the muon flux depends on the corresponding neutrino muon flux, and the differential neutrino spectrum is given by
\begin{equation}  \label{contributionsneutrinoflux}
\frac{dN_{\nu}}{dE_{\nu}}=\frac{\Gamma_a}{4\pi D^2}\sum_f B_{\widetilde N}^f  \frac{dN_{\nu}^{f}}{dE_{\nu}} \; ,
\end{equation}
see for example Ref.~\cite{DarkSUSY}. Appendix~\ref{appendix} contains a detailed discussion about the mass dependence of this equation.

The IceCube detector records the Cerenkov light from relativistic charged particles in its volume. Cosmic ray showers create a muon background signal that can be controlled by selecting for upward-going and contained muon events. The atmospheric neutrino background is well understood and may be subtracted away from the signal.

In addition to the muon flux through IceCube, electromagnetic and hadronic cascades inside the detector might also allow sneutrino dark matter detection. Electromagnetic cascades occur via charged current interactions. By depositing some of the incoming neutrino energy in taus and electrons, Bremsstrahlung radiation produces a localized cascade of energy that the digital optical modules of IceCube can record. In the results of Appendix~\ref{cascades} we have ignored any contribution from the charged current electromagnetic cascades of the muons, since their contribution has already been considered in the form of Cerenkov radiation from the muon tracks.
Hadronic cascades occur for both neutral current and charged current interactions. As the neutrino scatters off of a nucleus in the detector, the nucleus breaks up and produces products such as pions that in turn decay into detectable photons. Note that for neutral current interactions the energy of the outgoing neutrino is lost and is not recorded in any cascade.
The energy from localized electromagnetic and hadronic cascades is much harder to reconstruct compared to muon tracks but still might produce an interesting signal in the detector, see Appendix~\ref{cascades}.

\section{Model Results} \label{ModelResults}

The annihilation of sneutrinos in the Sun and Earth results in a flux of particle events through the IceCube detector that are calculated using DarkSUSY, which uses results from WimpSim \cite{DarkSUSY,WimpSim}. The calculations account for neutrinos produced via decays, as well as neutrino oscillation, loss via charged current interactions and scattering via neutral current interactions. DarkSUSY default parameters are used, which include a Gaussian dark matter velocity distribution and an NFW halo profile. Realistic Sun and Earth density profiles are integrated over numerically according to~\cite{Gould}. For both {\bf{case 1}} and {\bf{case 2}}, the maximum spin-independent cross section allowed by the $Z^{\prime}$ limits is used. 
Similarly, the annihilation cross section is fixed at $3 \times 10^{-26} \mathrm{cm^3/s}$ ($3 \times 10^{-23} \mathrm{cm^3/s}$) for {\bf{case 1}} ({\bf {case 2}}). Finally, the results presented in the subsections below use the convention of a detector energy threshold of 1 GeV. IceCube effective areas have not been calculated for our model, but we anticipate that they would be slightly larger than those used for the MSSM scenarios since we have a slightly harder spectrum. This is especially true in {\bf{case 1}}.

\subsection{Sensitivity to Neutrino Flavor}
\label{NeutrinoBranchingRatios}

\begin{figure}[h!]
\centering
\subfigure[~Sun Muon Neutrino Flux, {\bf {Case 1}}]{\label{fig:SunNeutrinosNoPamelaCompare}
\includegraphics[width=.42\textwidth]{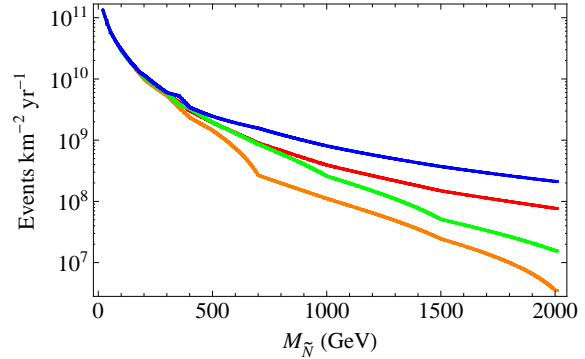}}
\subfigure[~Sun Muon Neutrino Flux, {\bf {Case 2}}]{\label{fig:SunNeutrinosPamelaCompare}
\includegraphics[width=.42\textwidth]{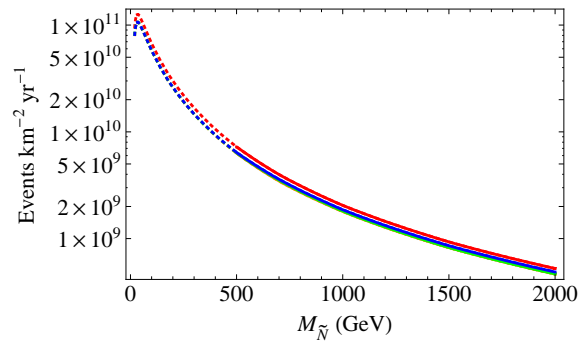}}
\caption[Sun Muon Neutrino Flux, Neutrino Branching Ratios]{\label{fig:SunCompare} Total muon neutrino rates received at the Earth for the $U(1)_{B-L}$ model as a function of the sneutrino mass in the case of sneutrino dark matter capture and annihilation in the Sun. The results are for one year of detection with IceCube. The $B-L$ model is robust to changes in the neutrino branching ratios. 100\% branching to $\nu_e$, $\nu_\mu$ and $\nu_\tau$ is shown in orange (bottom line in {\bf{case 1}}), green (second line from the bottom in {\bf{case 1}}) and blue (top line in {\bf{case 1}}) respectively. Results of equal branching to neutrino flavors are in red (second line from the top in {\bf{case 1}}).}
\end{figure}

For the results that follow we have considered equal branching to the three flavors of LH neutrinos, but in principle this need not be the case. The exact flavor composition of LH neutrinos produced from sneutrino annihilation in the Sun depends on the detailed structure of Majorana and Dirac couplings in the neutrino sector.
In Fig.~\ref{fig:SunCompare}, the resulting muon neutrino flux for a $100\%$ branching ratio to a single flavor is compared to equal flavor ratios in both {\bf case 1} and {\bf case 2} (upper and lower panels respectively).

It is seen from the upper panel that in {\bf {case 1}} for sneutrino masses below $300$ GeV (LH neutrino energy below 150 GeV) flavor composition of produced neutrinos does not matter since oscillations are very efficient at low energies and easily mix the neutrino flavors. Therefore 100\% $\nu_{e}$, $\nu_{\mu}$, or $\nu_{\tau}$ each leads to the same $\nu_{\mu}$ signal at the detector. However at high energies oscillation length $L_{\rm osc} \propto E_{\nu}/\Delta m^2$ elongates, and oscillations become less efficient. Here $\Delta m^2$ is the difference between $({\rm mass})^2$ of neutrino mass eigenstates. This effect is most important for $\nu_e$'s since they oscillate to $\nu_\mu$'s via the small mass splitting responsible for solar neutrino oscillations $\Delta m^2_{\rm sol}$. This is why the $\nu_\mu$ flux at the detector falls quickly for $100\%$ $\nu_e$ branching ratio at high energies.
The effect is less pronounced for $100\%$ $\nu_{\mu}$ and $\nu_\tau$ branching ratios because the relevant mass splitting is the one responsible for atmospheric neutrino oscillations $\Delta m^2_{\rm atm}$, which is much larger. However, it is seen that the $\nu_\mu$ flux for $100\%$ $\nu_\mu$ branching ratio is less than that for $100\%$ $\nu_\tau$ branching ratio at high energies. This is because of charged current interactions inside the Sun whose cross section is proportional to the neutrino energy. These interactions convert muon neutrinos to muons that are quickly stopped in the Sun due to electromagnetic interactions that result in attenuation of the neutrino flux. Charged current interactions also convert tau neutrinos to taus. However, due to their much shorter lifetime, they decay back to $\nu_\tau$ before any significant energy loss. Nevertheless, for sneutrino masses up to 1.5 TeV, the result for equal branching ratios to three flavors is within a factor of a few
  compared with the $100\%$ branching ratio to a single neutrino flavor. Moreover, for a typical model, it is unlikely that sneutrino annihilation produces only one flavor of RH neutrinos. Therefore equal branching to the three flavors is a good approximation in {\bf {case 1}}.

In {\bf case 2}, the lower panel\footnote{The effect of the 1 GeV conventional energy threshold in the spectrum can be seen at low masses as more of the neutrino signal is lost under the threshold; this causes the maximum event rate to move to the right from the edge of the graph. This effect is not evident in {\bf{case 1}} since the majority of the neutrino flux arrives at higher energies and is unaffected by the small threshold.}, there is virtually no difference between various flavor compositions. This is because sneutrino annihilation mainly produces taus in this case (the branching ratio for production of RH neutrinos is only 10\%). Hence equal branching to the three flavors is a nearly perfect approximation in this case.

We conclude that our results do not depend critically on the choice of neutrino flavor branching ratios in either case.

\subsection{Contributions to Muon Flux} \label{ChannelContributions}

It is worth emphasizing that {\bf {case 1}} and {\bf {case 2}} yield different neutrino signals. In {\bf {case 1}}, LH neutrinos are produced from two-body decay of (almost non-relativistic) RH neutrinos.
This produces a delta function in the energy of the LH neutrinos at one-half the mass of the sneutrino dark matter\footnote{There is one additional potential source for neutrinos: the Higgs produced from the decay of the RH neutrinos can itself decay to a $b \bar{b}$ pair. We checked that this contribution gives only a few percent change in the signal. We therefore neglect it in our numerical calculation for the sake of simplicity.}.
On the other hand, in {\bf case 2}, the sneutrino dominantly annihilates to $\phi \phi$ final states, and each $\phi$ decays to a fermion-antifermion pair via a one-loop diagram. The partial decay rate of $\phi$ is proportional to the squares of the mass of the resulting fermion and the fourth power of its $B-L$ charge~\cite{Allahverdi:2009ae,ADRS}. As a result, the largest contribution to the annihilation is from taus ($\approx 74 \% $) and bottom quarks ($\approx 16 \%$), where the quark signal is suppressed due to the $B-L$ charge. Both of these final states produce neutrinos via three-body decay that results in a spread in energy signal.

\begin{figure}[h!]
\centering
\subfigure[~Sun Muon Neutrino Flux]{\label{fig:SunNeutrinoChannelsNP}
\includegraphics[width=.42\textwidth]{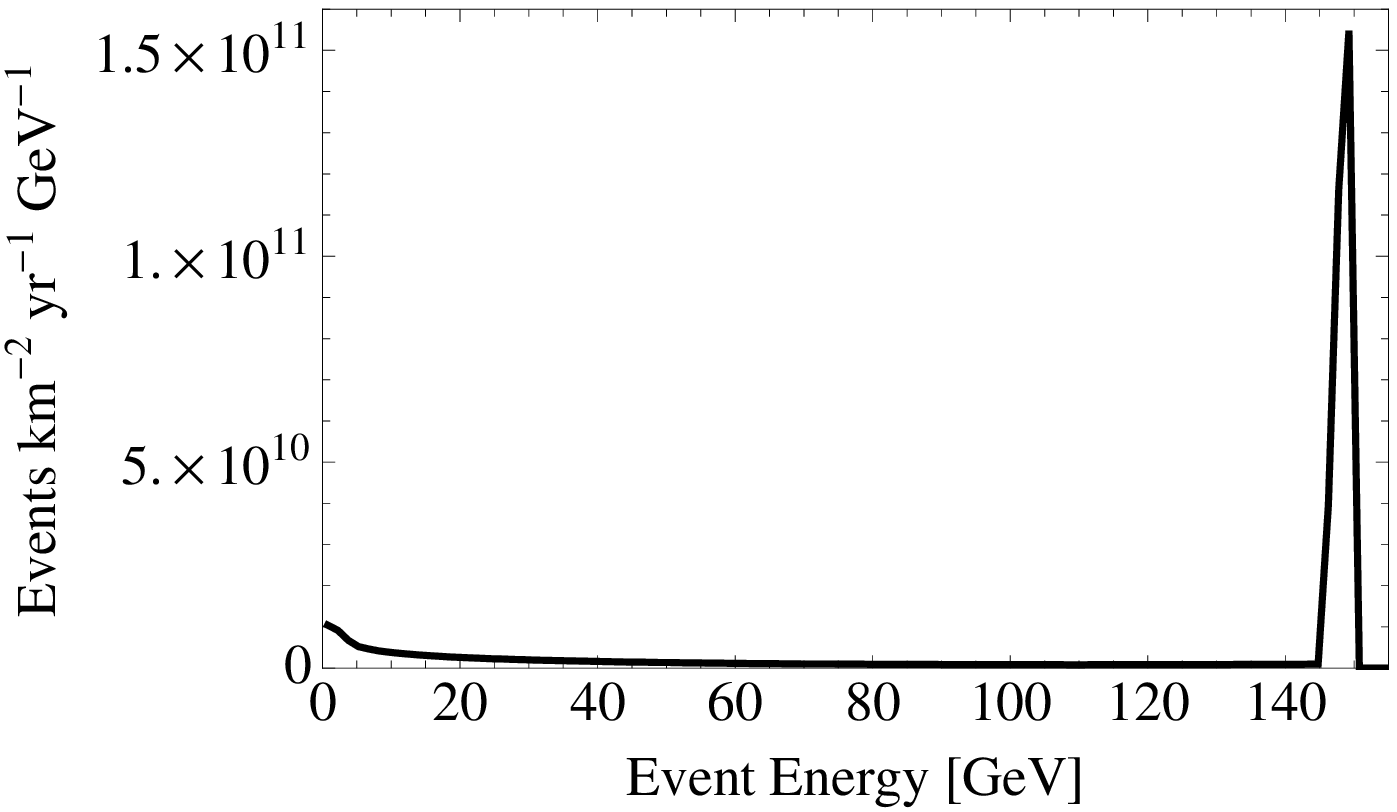}}
%
\subfigure[~Sun Muon Flux]{\label{fig:SunMuonChannelsNP}
\centering
\includegraphics[width=.42\textwidth]{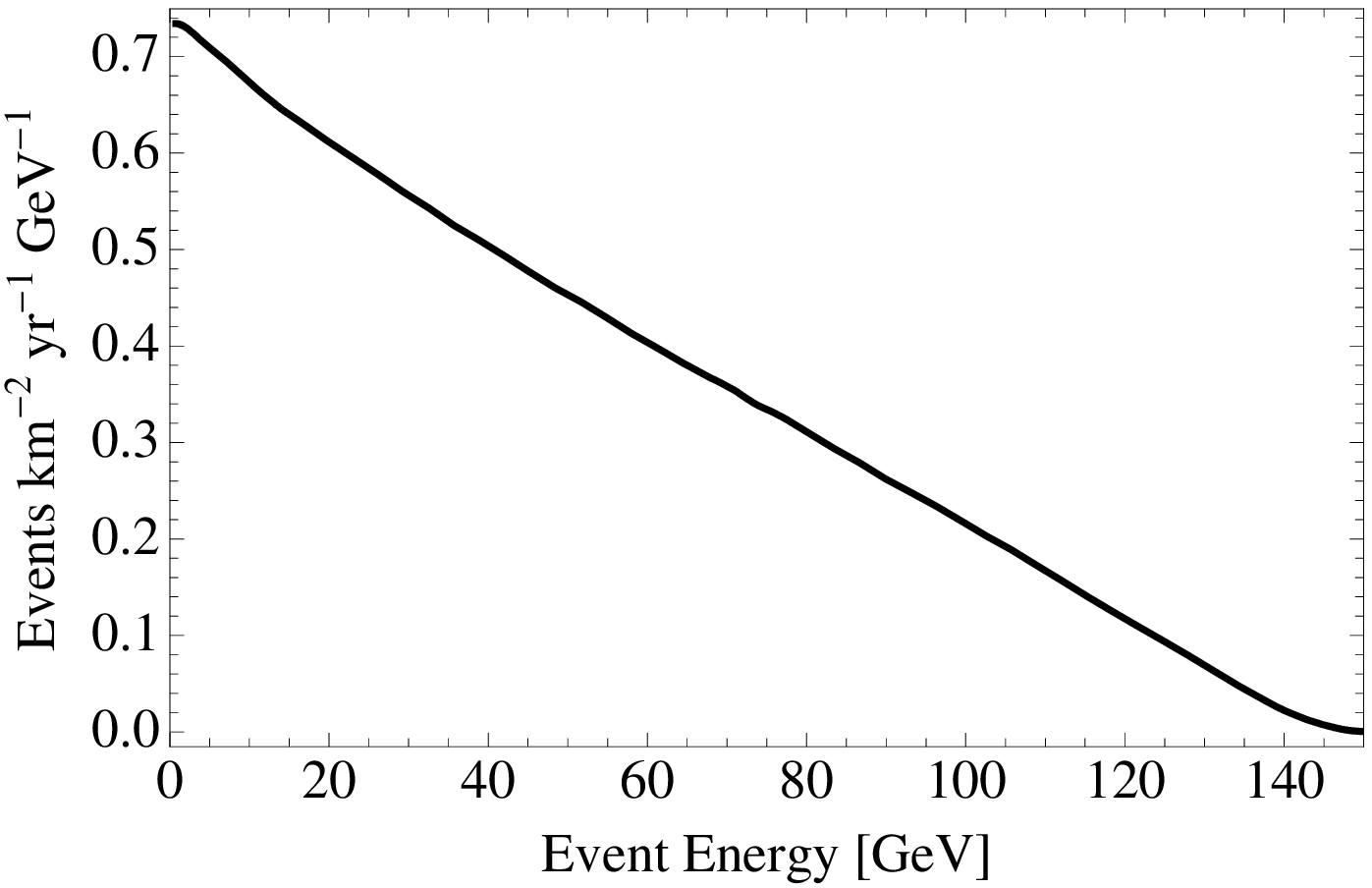}}
\caption[Sun Channels]{\label{fig:SunChannelsNP}In the upper (lower) panel, muon neutrino (muon) flux through IceCube from annihilation of 300 GeV sneutrinos in the Sun for {\bf {case 1}}.}
\end{figure}
%
Fig.~\ref{fig:SunNeutrinoChannelsNP} shows the muon neutrino flux energy spectrum through a kilometer squared of IceCube in one year for a 300 GeV sneutrino for {\bf {case 1}}. The delta function at half the mass of the sneutrino can be seen clearly. A small portion of muon neutrinos from this initial annihilation state are scattered via neutral current interactions inside the Sun to lower energies. This produces the slight bump in the spectrum at low energies. Fig.~\ref{fig:SunMuonChannelsNP} plots the resulting muon flux from the charged current interactions inside the IceCube detector. As expected for a monochromatic incident neutrino, the spectrum of muons has a linear dependence on energy.

\begin{figure}[h!]
\centering
\subfigure[~Sun Muon Neutrino Flux]{\label{fig:SunNeutrinoChannels}
\includegraphics[width=.42\textwidth]{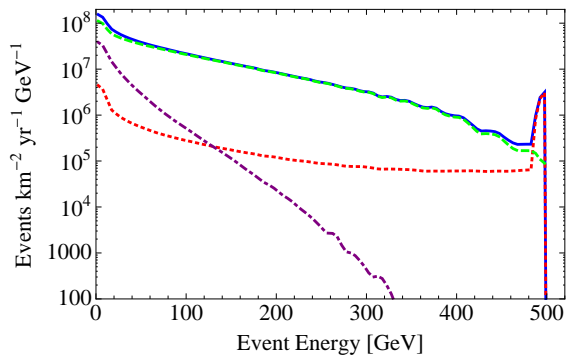}}
\subfigure[~Sun Muon Flux]{\label{fig:SunMuonChannels}
\centering
\includegraphics[width=.42\textwidth]{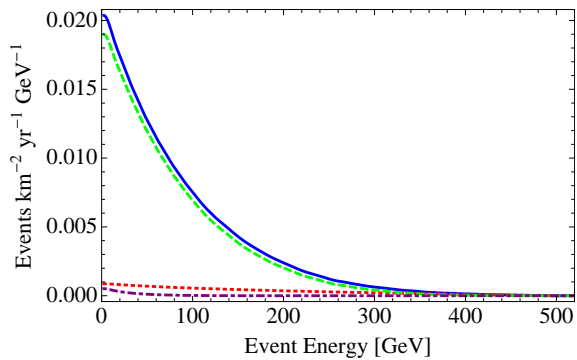}}
\caption[Sun Channels]{\label{fig:SunChannels}The same as Fig.~{\ref{fig:SunChannelsNP}}, but with a 1 TeV sneutrino in {\bf {case 2}}. Individual annihilation channels are shown: neutrino (red, dotted), tau (green, dashed), bottom quark (purple, dot-dashed) and all channels (blue, solid).
}
\end{figure}

For {\bf{case 2}}, the delta function from the neutrino channel at the detector is subdominant to the other annihilation channels, see Fig.~\ref{fig:SunNeutrinoChannels}. First, consider that the sneutrino annihilation mainly produces taus and bottom quarks that subsequently produce LH neutrinos via three-body decays. Second, due to the larger sneutrino mass of 1 TeV (in order to explain the PAMELA data), the LH neutrinos produced from two-body decays have a higher energy than in the 300 GeV case. Therefore they lose energy via neutral current interactions and get absorbed via charged current interactions inside the Sun more efficiently. As a result of both of these facts,
there are more neutrinos with low energies at the detector from each channel in this case than in {\bf{case 1}}. This also is reflected in the spectrum of muon flux, shown in Fig.~\ref{fig:SunMuonChannels}, which does not show a linear dependence on energy due to the presence of three-body decays. This is in contrast to Fig.~\ref{fig:SunMuonChannelsNP}.

The muon event signal from annihilation in the Earth for {\bf{case 1}} and {\bf{case 2}} is too small to detect since the dark matter population has not reached equilibrium; therefore, the production of neutrinos depends on both the scattering cross section and the annihilation cross section, which is small in this scenario. 
\subsection{Mass Dependence of Muon Flux} \label{Mass Dependence of Muon Flux}

\begin{figure}[h!]
\begin{center}
\includegraphics[width=7cm]{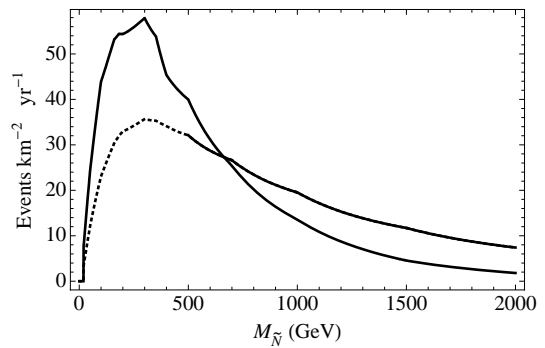}
\end{center}
\begin{center}
\parbox{8.0cm}{\caption[b]{\label{fig:MassPlotsSunMuons}Total muon rates detected at the Earth from annihilation of sneutrino dark matter in the Sun
as a function of the sneutrino mass.
The results are for one year of detection with IceCube. {\bf{Case 1}} ({\bf{case 2}}) is the highest (lowest) peaked line. The dotted line denotes the mass range where one cannot explain the PAMELA data using {\bf{case 2}} anymore.}}
\end{center}
\end{figure}

Fig.~\ref{fig:MassPlotsSunMuons} shows our results for the total muon rate integrated over energy as a function of
the sneutrino mass $m_{\widetilde N}$ for annihilation in the Sun\footnote{The apparent discrete nature of these plots occurs because only a few values of sneutrino mass are recorded in the WimpSim tables used by DarkSUSY; the program interpolates between these points. The effect is numerical and not physical.}. The figure shows both the {\bf{case 1}} and {\bf{case 2}} rates in events $\mathrm{km}^{-2}$ $\mathrm{yr}^{-1}$. The plots have two characteristics: an increase at lower masses culminating in a peak followed by a general decrease in event rates at higher masses.

The decrease of the event rates for higher $m_{\widetilde N}$ is reflective of the decrease of the neutrino flux due to the kinematic suppression of sneutrino capture (the factor scales approximately like $1/m_{\widetilde N}$ for large masses\footnote{See Appendix~\ref{appendix} for a more detailed definition of ``large''.}). The linear increase at low $m_{\widetilde N}$ is explained by the linear dependence of the cross section for charged current interactions on the energy of neutrinos at the detector (which is proportional to the sneutrino mass). The {\bf{case 1}} signal is larger than the {\bf{case 2}} signal for lower values of sneutrino mass. LH neutrinos are produced in two-body decays in {\bf {case 1}} versus three-body decays in {\bf {case 2}}, and hence have a higher energy. As a result, the cross section for conversion of neutrinos to muons at the detector is larger in {\bf {case 1}}. However, for large sneutrino masses {\bf {case 1}} has a smaller signal than {\bf 
 {case 2}}. The produced LH neutrinos, $100 \%$ of {\bf{case 1}} products, get absorbed via charged current interactions or lose energy via neutral current interactions inside the Sun more efficiently because of their larger energy, thus a smaller number of neutrinos arrive at the detector.

Refs.~\cite{C.DeClercq,Abbasi:2009uz} display sensitivity plots for the detection of a muon signal in the case of standard neutralino dark matter annihilation in the Sun and Earth respectively. In the case of the Earth, more than 12 events are needed for a DM mass between 70 GeV and 4 TeV. In the case of the Sun the number of events needed drops linearly as a function of mass starting from 300 events at 70 GeV down to 70 events at
300 GeV. Beyond 300 GeV up to 4 TeV, the number of events needed remains fixed at 70. This provides a hint that one could detect the event rates caused by sneutrinos despite some differences between the sneutrino and neutralino dark matter spectra.
These differences are due to unequal numbers and weighting of neutrino production channels, but the somewhat harder spectrum of the sneutrino model will make IceCube slightly more sensitive to the model. Hence, we can expect that it might be possible to detect muon neutrinos produced by sneutrino annihilation for sneutrino masses around 300 GeV for the Sun, cf.~Fig.~\ref{fig:MassPlotsSunMuons}. Note that a large range of masses would be accessible with only an order of magnitude improvement in sensitivity.

In summary, if the dark matter mass is determined from measurements at the LHC, then we can read the maximum number expected for the Sun muon rate in the $B-L$ model from Fig.~\ref{fig:MassPlotsSunMuons}\footnote{Since we have used the upper bound on the sneutrino-proton scattering cross section in our calculations, the number of muon events cannot be larger than that given in Fig.~{\ref{fig:MassPlotsSunMuons}}.}. Thus, for a known sneutrino mass, observation of a muon signal exceeding the number given in Fig.~\ref{fig:MassPlotsSunMuons} will rule out the $B-L$ model. The largest number of muon events from the Sun in the entire depicted mass range is 58 $\mathrm{km}^{-2}$ $\mathrm{yr}^{-1}$ (36 $\mathrm{km}^{-2}$ $\mathrm{yr}^{-1}$) for {\bf {case 1}} ({\bf {case 2}}). Therefore detection of a muon signal larger than this will rule out the $B-L$ model regardless of the sneutrino mass.

In the case of the Earth, as mentioned in the previous subsection, there is no prospect for a potential detection at IceCube for the standard halo model. The number of muon events is 6 orders of magnitude below the minimum measurable Earth rate of 12 $\mathrm{km}^{-2}$ $\mathrm{yr}^{-1}$ in this case.

\section{Dark Matter Disc in the Milky Way} \label{DM Disc in the Milky Way}

In our analysis, so far we have assumed a Gaussian like velocity distribution for dark matter particles
with a typical value for the three dimensional velocity dispersion of $\sigma_v = 270$ $\mathrm{km}$ $\mathrm{sec}^{-1}$ and $\vert v_{Sun}\vert=220$ $\mathrm{km}$ $\mathrm{sec}^{-1}$ for the velocity of the solar system with respect to the halo. However, there are recent speculations about the existence of a dark matter thick disc in the Milky Way in addition to the baryonic one, see e.~g.~\cite{Read:2008fh,Read:2009iv}. This dark matter disc is caused by the accretion of Milky Way satellite galaxies and their corresponding baryonic and dark matter. As dynamical friction causes the satellite galaxies to accrete onto the disc, tidal forces disrupt the satellites~\cite{Read:2009iv}. Galaxy formation simulations find the density of the dark matter disc $\rho_{\rm dark}$ to be in the range $\approx 0.25-1.5$ times the local halo dark matter density $\rho_{\rm halo}$~\cite{Read:2009iv}.Possible ranges for the solar system velocity and velocity dispersion of the dark matter disc are$\vert v_{Sun}\vert\approx 0-150$ $\mathrm{km}$ $\mathrm{sec}^{-1}$ and $\sigma_v \approx 87-1
 56$ $\mathrm{km}$ $\mathrm{sec}^{-1}$.

\begin{figure}[h!]
\centering
\subfigure[~Earth muons, {\bf{Case 1}}]{\label{EarthVelocityNoPamela}
\includegraphics[width=.42\textwidth]{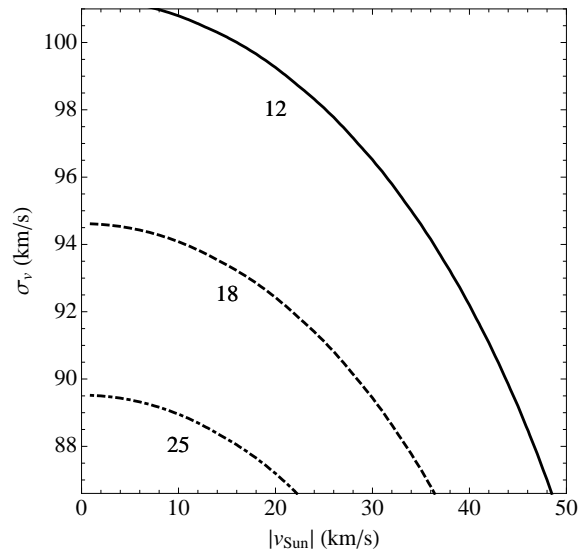}}
\subfigure[~Earth muons, {\bf{Case 2}}]{\label{EarthVelocityPamela}
\includegraphics[width=.42\textwidth]{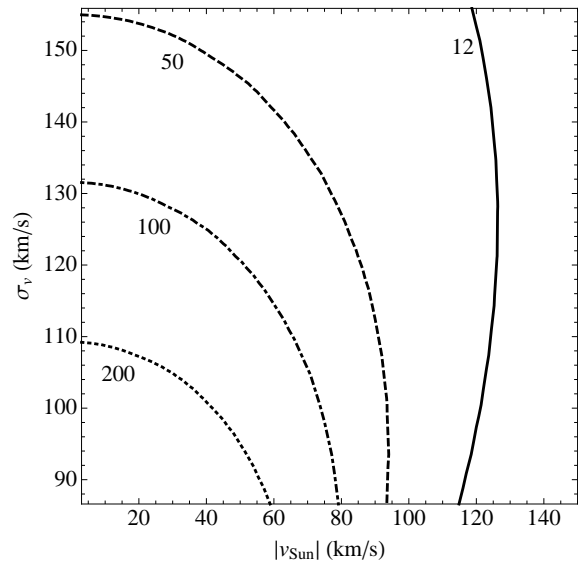}}
\caption[Earth Velocity Changes]{\label{fig:EarthVelocityVaryVel} Total Earth-annihilation muon event rates inside the detector per kilometer squared per year for a 300 GeV (in {\bf {case 1}}) and 1000 GeV (in {\bf{case 2}}) sneutrino.}
\end{figure}

Fig.~\ref{fig:EarthVelocityVaryVel} shows the Earth muon rate when we scan about the relevant parameter space for the allowed values of $\vert v_{Sun}\vert$ and $\sigma_v$ in {\bf{case 1}} and {\bf{case 2}}. We used the fixed ratio $\rho_{\rm dark}/\rho_{\rm halo}=1$. {\bf {Case 2}} has a sufficient total event rate ($\geq$ 12 $\mathrm{km}^{-2}$ $\mathrm{yr}^{-1}$) for nearly the whole allowed parameter space.
The constraint of the parameter space is more pronounced for {\bf{case 1}}. The allowed combinations are roughly given by a triangle with maximal values of  $\vert v_{Sun} \vert = 47 \rm{km}/s$ and $\sigma_{v} = 100 \rm{km}/s$. The differences in the allowed parameter space for the two cases reflects the fact that the Earth is not in equilibrium yet. Thus the muon neutrino signal and the corresponding muon flux still depends on the annihilation cross section, which is three orders of magnitude larger for {\bf{case 2}}. However, we see that in both cases the Earth rates have increased to detectable rates, several orders of magnitude higher than the standard halo model that has higher $\vert v_{Sun} \vert $ and $\sigma_{v}$, used in the previous section\footnote{The usage of a free space Gaussian velocity distribution means that our calculated event rates are an upper bound. There are many proposed parameterizations for the dark matter velocity distribution, and a Gaussian distribution belongs to the scenarios with the highest resulting event rates, see~\cite{Bruch:2009rp}.}.

\begin{figure}[h!]
\centering
\includegraphics[width=.42\textwidth]{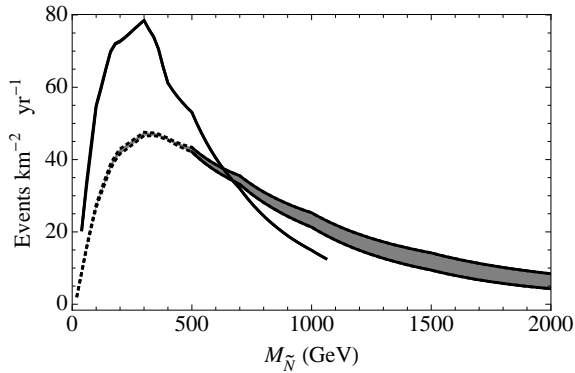}
\caption[Earth Channels]{\label{fig:BandSunRates} Total Sun-annihilation muon rates inside the detector for the sneutrino dark matter with modified velocity distributions that yield 
Earth-annihilation rates of at least 12 events per year per $\mathrm{km}^2$. Upper (lower) curve shows the {\bf{case 1}} ({\bf{case 2}}). The dotted lines denote the mass range where one cannot explain the PAMELA data using {\bf{case 2}} anymore.
}
\end{figure}

A change in the velocities and dispersions also modifies the corresponding total Sun event rates. This is comparatively modest for neutralino dark matter where it is at most one order of magnitude, see~\cite{Bruch:2009rp}. Fig.~\ref{fig:BandSunRates} shows a band of allowed total Sun muon rates for the sneutrino dark matter. These rates are given again as a function of the sneutrino mass and under the requirement that we have a measurable Earth rate of at least 12 events $\mathrm{km}^{-2}$ $\mathrm{yr}^{-1}$. Any variation of the event numbers for a fixed mass arises as a result of the use of velocities $\vert v_{Sun}\vert$ and dispersions $\sigma_{v}$ within the required parameter ranges of Fig.~\ref{EarthVelocityNoPamela} and~\ref{EarthVelocityPamela}. A comparison between Fig.~\ref{fig:MassPlotsSunMuons} and~\ref{fig:BandSunRates} shows an increase of $\approx 30\%$ in the Sun muon rates for the sneutrino dark matter.

The band of total muon rates for {\bf {case 1}} is noticeably thinner than for {\bf{case 2}}. It is seen from Figs.~\ref{EarthVelocityNoPamela} and~\ref{EarthVelocityPamela} that {\bf{case 1}} has a smaller allowed parameter range with more than 12 events $\mathrm{km}^{-2}$ $\mathrm{yr}^{-1}$. Thus the corresponding ratio between the minimal and maximal value within the allowed range is much smaller than that in {\bf{case 2}}, and the possible change in the total Sun rates is comparatively small. Even for {\bf{case 2}} the differences between the highest and lowest rates for a fixed mass is about 40\% or less.

To summarize, a modified velocity distribution can substantially enhance the Earth muon rate for the sneutrino dark matter  beyond the detection threshold of 12 $\mathrm{km}^{-2}$ $\mathrm{yr}^{-1}$. It also raises the maximum Sun muon rate to 78 events $\mathrm{km}^{-2}$ $\mathrm{yr}^{-1}$ (48  $\mathrm{km}^{-2}$ $\mathrm{yr}^{-1}$) in {\bf {case 1}} ({\bf {case 2}}). Observation of the Sun muon rates larger than these will rule out the $B-L$ model regardless of the sneutrino mass or Earth rates.

%
%
\section{Comparison with mSUGRA} \label{Comparison}

Minimal supergravity (mSUGRA) is a constrained version of the MSSM that depends only on four parameters and one sign. These are $m_0$ (the universal soft breaking mass at the grand unification scale), $m_{1/2}$ (the universal gaugino soft breaking mass at the grand unification scale), $A$ (the universal trilinear soft breaking mass at the grand unification scale), ${\rm tan} \; \beta$ (the ratio of MSSM Higgs VEVs at the electroweak scale) and the sign of $\mu$ (the MSSM Higgs mixing parameter). The mSUGRA dark matter candidate is the lightest neutralino.

The parameter space of the mSUGRA model has three distinct regions allowed by the dark matter constraints~\cite{darkrv}: (i) the co-annihilation region where both $m_0$ and  $m_{1/2}$ can be small, (ii) the hyperbolic branch/focus point region where the dark matter has a large Higgsino component and $m_0$ is very large
but $m_{1/2}$ is small, and (iii) the funnel region where both $m_0$ and $m_{1/2}$ are large and the dark matter annihilation occurs through heavy Higgs bosons in the $s$-channel. We note that a bulk region (where none of the above properties hold) is now almost ruled out due to other experimental constraints. Among these three regions, the neutralino has a large capture rate in the hyperbolic branch/focus point region due to a large Higgsino component that results in a large spin-dependent scattering cross section via $Z$ exchange. In this section we compare mSUGRA hyperbolic branch/focus point scenarios with the $B-L$ model.

\begin{figure}[h!]
\begin{center}
\includegraphics[width=7cm]{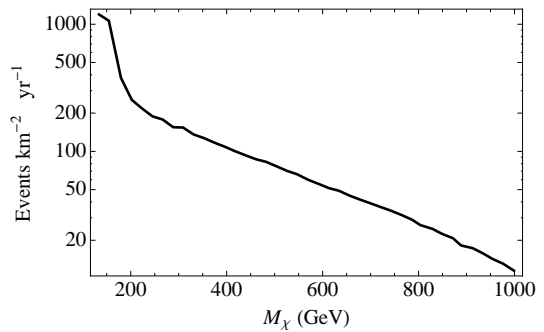}
\end{center}
\begin{center}
\parbox{8.0cm}{\caption[b]{\label{fig:MassPlotSunRateFocusPoints}Total Sun-annihilation muon rates inside the detector for mSUGRA hyperbolic branch/focus point scenarios as a function of the neutralino mass.
The results are for one year of detection with IceCube.}}
\end{center}
\end{figure}

Fig.~\ref{fig:MassPlotSunRateFocusPoints} shows the total Sun muon rate as a function of the neutralino mass for mSUGRA hyperbolic branch/focus points. A comparison with Fig.~\ref{fig:MassPlotsSunMuons} shows that these scenarios always have a higher total muon rate in the plotted mass range than the $B-L$ model. The hyperbolic branch/focus point models yield larger muon rates by between more than one order of magnitude and a factor of $1.5$ for dark matter masses in the $100-800$ GeV range.
Even for masses up to 400 GeV the hyperbolic branch/focus point scenarios provide rates higher than 100 events $\mathrm{km}^{-2}$ $\mathrm{yr}^{-1}$. These higher rates are explained by the bigger spin-dependent scattering cross sections, which are a few orders of magnitude larger than the upper bound on the spin-independent cross section for the sneutrino dark matter. The spin-dependent scattering cross section for the $B-L$ model is zero because $U(1)_{B-L}$ is a vectorial symmetry. Since the Sun mainly consists of hydrogen, the spin-dependent piece contributes dominantly for the mSUGRA case.

However, it is interesting that despite having a much smaller scattering cross section, the $B-L$ model can yield muon rates that are roughly comparable to the mSUGRA scenarios. Sneutrino annihilation dominantly produces leptons, i.e., RH neutrinos in {\bf {case 1}} and taus in {\bf {case 2}}, which subsequently decay to LH neutrinos $100\%$. On the other hand, neutralino annihilation in the hyperbolic branch/focus point scenarios dominantly produces quark final states that have a small branching ratio for decay to neutrinos.

Furthermore, despite lower event rates, sneutrino dark matter still produces a distinctive linear spectrum in the muon flux. As illustrated in subsection~\ref{ChannelContributions}, this feature is caused by the delta function in energy for the neutrino spectrum and can be used to distinguish between the $B-L$ model and the hyperbolic branch/focus point scenarios as long as energy binning of the differential muon rate with respect to the energy is precise enough at IceCube.

\begin{figure}[h!]
\begin{center}
\includegraphics[width=7cm]{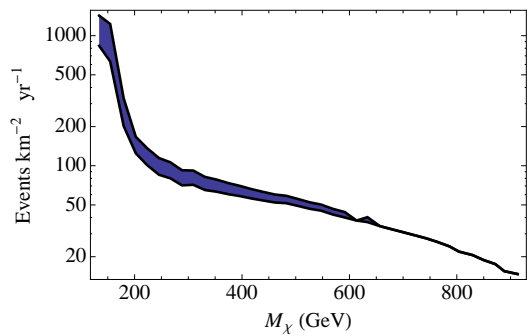}
\end{center}
\begin{center}
\parbox{8.0cm}{\caption[b]{\label{fig:MassPlotSunMuonsFocusPoints} Total muon rates detected inside the Earth for mSUGRA Focus point scenarios as a function of the neutralino mass in the case of neutralino DM capture and annihilation in the Sun. Rates for a range of velocities and dispersions for which the corresponding Earth rates are at least 12 events per year per $\mathrm{km}^2$ are shown in the shaded region. The results are for one year of detection with IceCube.}}
\end{center}
\end{figure}

Fig.~\ref{fig:MassPlotSunMuonsFocusPoints} shows the counterpart of Fig.~\ref{fig:BandSunRates} for mSUGRA hyperbolic branch/focus point scenarios. The range of velocities and dispersions for which the corresponding Earth rates are at least 12 events $\mathrm{km}^{-2}$ $\mathrm{yr}^{-1}$ yields a band for the total Sun muon rates. We see that the range between the highest and lowest rates for a fixed mass does not exceed a factor of two even for masses below 200 GeV.

A scan about the whole parameter space of the modified velocity distribution yields a maximum of 13 events $\mathrm{km}^{-2}$ $\mathrm{yr}^{-1}$ from the Sun for a 1000 GeV neutralino in the hyperbolic branch/focus point sneutrino. The $B-L$ model with sneutrino masses of 1000 GeV and 1500 GeV gives rise to maximum values of 18 and 6 (25 and 14) events  $\mathrm{km}^{-2}$ $\mathrm{yr}^{-1}$ for {\bf{case 1}} ({\bf{case 2}}). In contrast, for a dark matter mass of 300 GeV, the maximum events $\mathrm{km}^{-2}$ $\mathrm{yr}^{-1}$ are 158 (hyperbolic branch/focus point), 79 ({{\bf{case 1}}) and 48 ({{\bf{case 2}}). Thus the hyperbolic branch/focus point rates are larger than the $B-L$ rates for low masses, but both are in the detectable range at IceCube. At high masses it becomes more difficult to distinguish between the hyperbolic branch/focus point and the $B-L$ models using maximal Sun rates; we would have to depend instead on the spectral features mentioned in Section~\ref{ChannelContributions}.

In the stau co-annihilation and Higgs resonance regions the lightest neutralino has a high gaugino fraction and therefore a much smaller spin-dependent cross section
that leads to much lower event rates than the $B-L$ model.
For example, even if we assume a modified velocity distribution without any minimal Earth event rate condition the maximum total Sun rate is less than 1 event $\mathrm{km}^{-2}$ $\mathrm{yr}^{-1}$ for a 300 GeV neutralino (compared with the maximum Sun rate of 158 events $\mathrm{km}^{-2}$ $\mathrm{yr}^{-1}$ for a hyperbolic branch/focus point scenario with the same mass). This is far below any detection threshold.

It is also important to note that the hyperbolic branch/focus point in the mSUGRA model is incompatible with the $g-2$ data, where there exists a 3$\sigma$ deviation from the SM value if the $e^+ e^-$ data is used to calculate the leading order hadronic contribution~\cite{g-2}. In the context of the $B-L$ model, {\bf {case 2}}, which can address the PAMELA puzzle, also becomes incompatible with $g-2$ data, however the generic $B-L$ model, i.e. {\bf{case 1}}, is still compatible.

\section{Conclusion}

We have considered prospects of indirect detection of the RH sneutrino dark matter in a $U(1)_{B-L}$ extension of the MSSM at the IceCube neutrino telescope. The sneutrinos captured in the Sun and Earth dominantly annihilate through $S$-wave processes at the present time. In a generic situation (called {\bf {case 1}}) the sneutrinos annihilate to RH neutrinos (annihilation cross section of $ 3 \times 10^{-26}$ cm$^3$/sec) that quickly decay to a LH neutrino and the MSSM Higgs. If one seeks an explanation for the recently observed positron excess from the PAMELA data (called {\bf {case 2}}), the sneutrinos with a mass $\geq 1$ TeV dominantly annihilate to the lightest Higgs in the $B-L$ sector (with an enhanced annihilation cross section of $3 \times 10^{-23}$ cm$^3$/sec) that rapidly decay to fermion-antifermion pairs (74\% taus, 16\% bottom quarks, and 10\% RH neutrinos). LH neutrinos are produced mainly from the three-body decay of taus. The muon neutrinos from sneutrino annihilation are converted to muons via charged current interactions at IceCube.

In both of the cases, sneutrino capture and annihilation inside the Sun reaches equilibrium. Consequently, the flux of neutrinos from the Sun is governed by the cross section for sneutrino-proton elastic scattering, which has an upper bound of $8 \times 10^{-9}$ pb from the LEP and Tevatron limits on the $Z^{\prime}$ mass (due to the vectorial nature of the $B-L$ symmetry, there is no spin-dependent piece). In Fig.~\ref{fig:MassPlotsSunMuons} we have shown the number of Sun muon events at IceCube as a function of the sneutrino mass for {\bf {case 1}} and {\bf {case 2}} (using the upper bound on the sneutrino-proton scattering cross section). In both cases, the number of events are potentially detectable by IceCube due to a harder neutrino spectrum. Thus once the dark matter mass is found from measurements at the LHC, observation of muon events larger than that given in Fig.~\ref{fig:MassPlotsSunMuons} will rule out the $B-L$ model.

For the standard halo model the capture and annihilation of sneutrinos inside the Earth does not reach equilibrium for either {\bf {case 1}} or {\bf {case 2}}, resulting in an event rate that is too small to be detected at IceCube. However, modified velocity distributions within the range allowed by recent simulations of the galaxy can lead to a substantially larger rate that exceeds the IceCube detection threshold of 12 $\mathrm{km}^{-2}$ $\mathrm{yr}^{-1}$ for events from annihilation in the Earth. Nevertheless, the Sun-annihilation muon rate can at most increase by 30\% for a modified velocity distribution, as shown in Fig.~\ref{fig:BandSunRates}. This implies that observation of a muon event rate larger than roughly 100 $\mathrm{km}^{-2}$ $\mathrm{yr}^{-1}$ from the Sun will all but rule out the $B-L$ model regardless of the dark matter mass.

We compared predictions of the sneutrino dark matter in the $B-L$ model with that of the neutralino dark matter in the mSUGRA model. Only hyperbolic branch/focus point scenarios in mSUGRA, which have a Higgsino type dark matter candidate and thus large spin-dependent contributions to the neutralino-proton elastic scattering cross section, give rise to Sun muon event rates that can be detected at IceCube. Even though scattering cross sections can be two to three orders of magnitude larger than the $B-L$ case, the muon rates do not scale directly with the cross section. This is because sneutrinos mainly annihilate into lepton final states (by virtue of the $B-L$ symmetry) that decay to neutrinos with 100\% efficiency, while neutralino annihilation dominantly produces quark final states that have a small branching ratio for decay to neutrinos. Moreover, the linear dependence of the muon spectrum on the energy in the case of the sneutrino dark matter (particularly {\bf {case 1}})
 , a common feature for neutrinos produced from the two-body decays, can be used to distinguish between the $B-L$ model and the hyperbolic branch/focus point scenarios. This will be feasible by a sufficiently precise energy binning of the differential muon rate at IceCube.

\begin{appendix}

\section{Mass Dependence of $\Gamma_A$} \label{appendix}

We analyze in detail here the contribution from Eq.~(\ref{annihilationeq}) to the mass behavior of Fig.~\ref{fig:MassPlotsSunMuons}. Any mass dependence from $dN_{\nu}^{f}/dE_{\nu}$ is ultimately washed out of the muon signal by the linear dependence of $\sigma_{CC, NC}$ on the neutrino energy, which dominates at low energy. The distance $D$ between the detector and the source and the branching fraction $B_{\widetilde N}^f$ into the final state $f$ are independent of $m_{\widetilde N}$. Thus, the annihilation rate at high energies is governed by the dependence of $\Gamma_A$ on the kinematic suppression factor. Eq.~(\ref{annihilationeq}) shows that this annihilation rate is proportional to the capture rate $C$. Ref.~\cite{Jungman:1995df} provides a parameterization of $C$ as a function of the energy:
\begin{eqnarray}
C(m_{\widetilde N})\propto\sum_i F_i(m_{\widetilde N}) S(m_{\widetilde N}/m_{N_i}) \sigma_i(m_{\widetilde N}) \ ,   \label{approxCR}
\end{eqnarray}
where the sum runs over all species $i$ of nuclei in the Sun or Earth, $F_i$ are the corresponding form factors,
$S$ is the kinematic suppression factor for capture of a sneutrino and the $\sigma_i$ are the individual
scalar cross sections for scattering from nucleus $i$. The effect of the $F_i$ dependence on mass is negligible because most of these form factors vary little from unity. Furthermore, $\sigma_i$ is not dependent on the sneutrino mass in the $B-L$ model since we have chosen a constant sneutrino-proton scattering cross section of $8\times10^{-9}$ pb (the upper bound implied by the $Z^{\prime}$ mass limits).
Thus the overall shape of the curves in Fig.~\ref{fig:MassPlotsSunMuons} can be understood by looking at $S(m_{\widetilde{N}})$.

$S$ can be parameterized by
\begin{eqnarray}
S(x)&=&\Big[\frac{A(x)^{1.5}}{1+A(x)^{1.5}}\Big]^{3/2} \ , \label{Sfunc}  \\
A(x)&=&1.5\frac{x}{(x-1)^2}\Big(\frac{<\!v_{esc}\!>^2}{\bar v^2}\Big) \label{Afunc} \ ,
\end{eqnarray}
where $\bar v=270$ $\mathrm{km}$ $\mathrm{sec}^{-1}$ is the velocity dispersion of the dark matter particles and  $<\!v_{\mathrm{esc}}\!>$ is the escape velocity of $1156$ $\mathrm{km}$ $\mathrm{sec}^{-1}$ and $13.2$
$\mathrm{km}$ $\mathrm{sec}^{-1}$ for the Sun and Earth respectively. $S(x)$ is bounded between zero and one. Moreover, it scales like  $1.5(<\!v_{\mathrm{esc}}\!>^2/{\bar v^2})/x$ for $x\rightarrow\infty$, and it peaks at one for $x=1$. Therefore, the exact location of the peak for each scattering element $i$ is determined by its corresponding nucleus mass $m_{N_i}$.
As mentioned in subsection~\ref{Mass Dependence of Muon Flux}, $S$ scales approximately like $1/m_{\widetilde N}$ for large masses. The meaning of large in this context depends on the value of the $(<\!v_{\mathrm{esc}}\!>/\bar v)^2$ ratio in comparison to $x$. For example, a value of $m_{\widetilde N}$ with $m_{\widetilde N}/m_{N_i}>(<\!v_{\mathrm{esc}}\!>/\bar v)^2$ is considered large.

\section{Cascade Signal} \label{cascades}

\begin{figure}[h!]
\centering
\subfigure[~Sun Cascades, {\bf{Case 1}} 300 GeV sneutrino]{\label{fig:SunCascadesNoPamela}
\includegraphics[width=.42\textwidth]{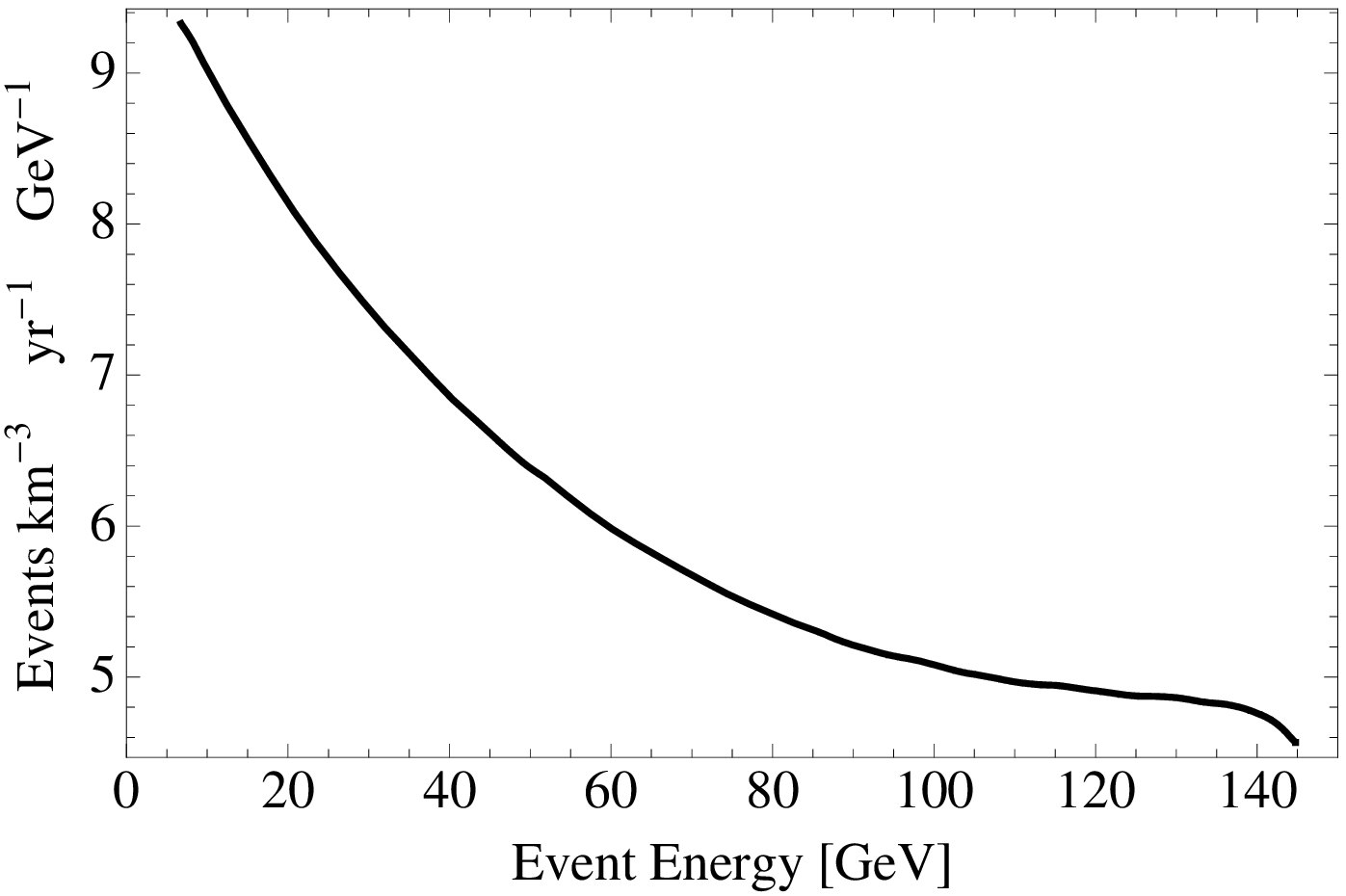}}
\subfigure[~Sun Cascades, {\bf{Case 2}} 1000 GeV sneutrino]{\label{fig:SunCascadesPamela}
\includegraphics[width=.42\textwidth]{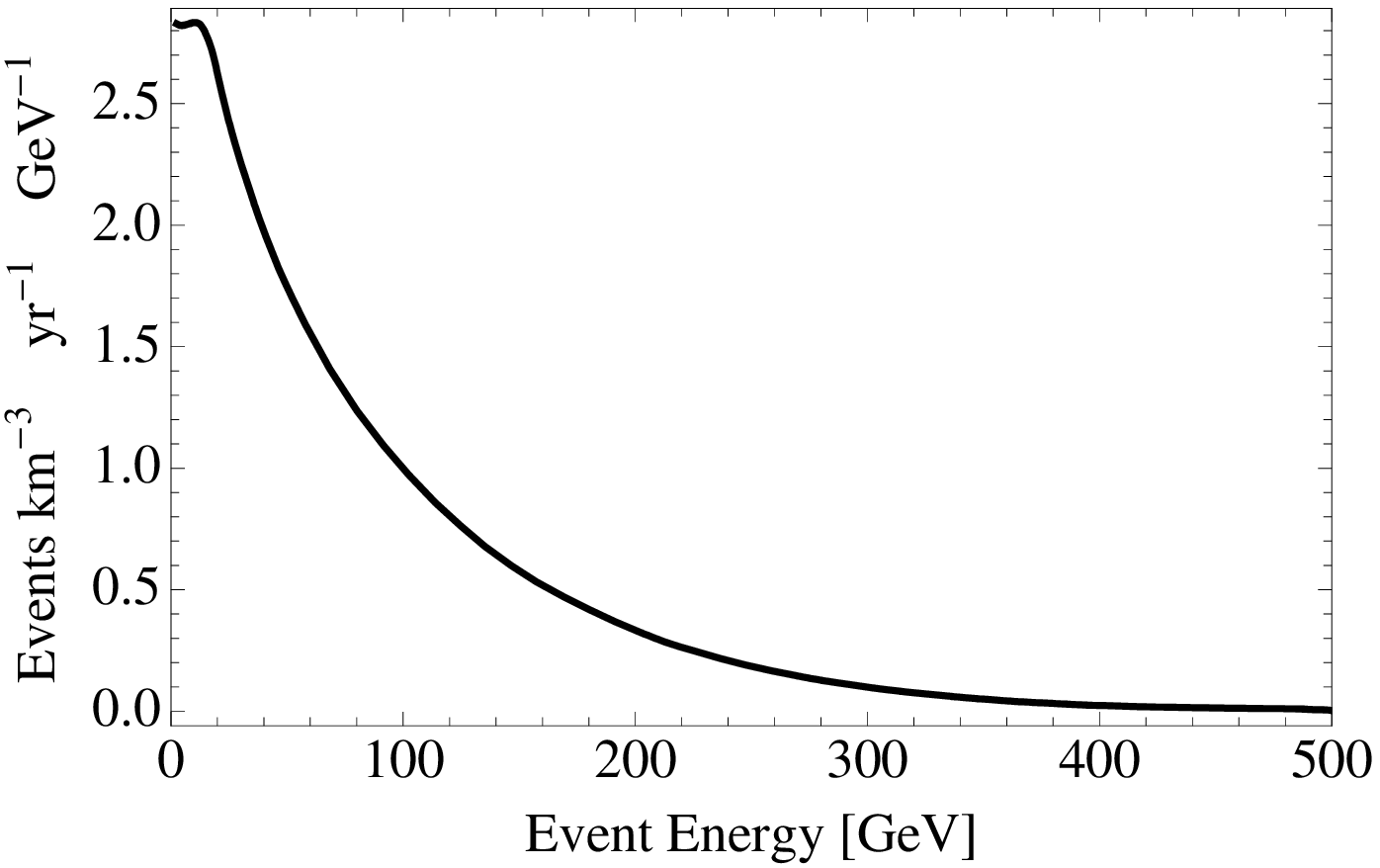}}
\caption[Sun Cascades]{\label{fig:SunCascades} Total electromagnetic and hadronic cascades inside the detector volume from sneutrino annihilation in the Sun.}
\end{figure}
Fig.~\ref{fig:SunCascades} plots the total energy spectrum from all cascades, both hadronic and electromagnetic (excluding the electromagnetic muon signal\footnote{The electromagnetic cascade from a muon signal is excluded from the graph since it is accompanied by a more discernable muon track, the subject of the body of this paper.}), per kilometer squared of detector per year for {\bf{case 1}} and {\bf{case 2}}. The general downward trend of the plot occurs because the hadronic signal dominates as it is produced by both charged current and neutral current interactions, while the upward-trending electromagnetic signal only receives contributions from the charged current interactions and excludes the muon signal altogether. The cross sections for hadronic processes decrease as the transferred energy to the nucleus goes up, hence creating the decreasing trend (high energy in the hadronic cascades corresponds to low energies in the electromagnetic cascades). We note that in the lower panel of the figure ({\bf case 2}) the cascade signal is depleted at high energies. This is because the produced neutrinos have higher energies (as a result of the higher sneutrino mass in this case), and therefore absorption and scattering effects inside the Sun are more important. This explains why the signal in {\bf{case 2}} is more steeply curved than the {\bf{case 1}} signal.

It is important to remember that it is not clear at this time whether IceCube will be able to distinguish between electromagnetic and hadronic cascades. As a result, while a single charged current interaction will result in both a hadronic and electromagnetic cascade, these may be recorded as a single event with the total energy of the incoming neutrino. Meanwhile, the hadronic cascade of neutral current events would be recorded correctly as a single event with only part of the energy of the incoming neutrino. While we have assumed in the above that individual cascade signals are separable this may not reflect experimental reality.
\end{appendix}

\section{Acknowledgement}

The authors wish to thank Spencer Klein and Carsten Rott for valuable discussions. The work of BD is supported in part by DOE grant DE-FG02-95ER40917.



\end{document}